\documentclass[9pt,twocolumn,twoside]{osajnl}

\journal{josaa} 

\title{ Theoretical and numerical investigation of internal conical refraction of structured light beams}

\author[1,*]{S. F. Caballero-Ben\'\i tez }
\author[2]{S. Hacyan}

\affil[1]{Instituto de F\'{\i}sica, LSCSC-LANMAC, Universidad Nacional Aut\'onoma de M\'exico, CP. 04510, Ciudad de M\'exico, M\'exico}
\affil[2]{Instituto de F\'{\i}sica , Universidad Nacional Aut\'onoma de M\'exico, CP. 04510, Ciudad de M\'exico, M\'exico}

\affil[*]{Corresponding author: scaballero@fisica.unam.mx}

\begin{abstract}
We present an ab-initio numerical investigation of the internal conical refraction of structured light beams in a biaxial crystal. Starting from the solutions of the  Fresnel equation, a theoretical analysis is developed without assuming any analytical approximation, thus obtaining a set of exact equations that can be solved by standard methods of integration for any impinging light beam. As examples of applications, we consider the particular cases of linearly and circularly polarized Gaussian and  Bessel beams inside a KTP crystal. The numerical calculations follow the evolution of the refracted beam inside the crystal. It is seen that for realistic boundary conditions, a refraction cone  appears  in a certain range of distances within the crystal and its shape is rather sensitive to the initial conditions.
\end{abstract}

\begin{document}

\maketitle
\section{Introduction}

The conical refraction of light in a biaxial crystal is an interesting phenomena that has attracted much attention since it was theoretically predicted, almost two centuries ago, by Hamilton \cite{H} and confirmed experimentally shortly afterwards   by Lloyd \cite{L} and Poggendorff \cite{P}. In more recent times, many works have been devoted to both the theoretical \cite{pb,lalor1,lalor2,warnick,belsstep,B2,B1,matos,khilo,turp1,turp2,turp3} as well as the experimental aspects of this phenomena \cite{R1,R2,pemi,ktp,nature}. The basic problem, however, is that the equations describing the phenomena are quite cumbersome and cannot be solved without assuming some approximations. For instance, the wave-vectors inside the crystal are given by the Fresnel equation, which is of fourth-order, and although analytic solutions can be obtained \cite{khilo}, they are too cumbersome to be used and most authors prefer to resort to some kind of approximations for the two modes that appear due to the anisotropic refraction. This latter approximation, in particular, could lead to the production of higher order Bessel beams from a lower order one \cite{chc,King,belruso,bels}.

In order to elucidate the validity of the approximate methods used so far, we present in this paper an ab-initio calculation of the evolution of a structured light beam of arbitrary shape along the optical axis inside a biaxial crystal. Our purpose is to obtain a set of equations given in such a form that they can be solved by standard methods of numerical calculation. Thus, we are able to  follow  the evolution of a realistic light beam inside the crystal, and visualize the formation of the refraction cone.

The plan of the paper is the following. In Section 2, we present, for the sake of completeness, the basic equations for conical refraction and work out all the relevant formulas. In Section 3, the boundary conditions are used to obtain the complete set of equations to be solved numerically. Section 4 is devoted to two important applications of our formalism: we study the evolution of circularly and linearly polarized Gauss and Bessel beams inside a KTP crystal. The results are presented in a series of graph obtained by numerical calculation in which the evolution of the refraction cone is clearly seen at various depths inside the crystal. In Section 5, we outline our procedure for the numerical simulations. In Section 6, we present some brief conclusions of our study.

\section{Geometry}

Maxwell's equations imply for the electric field ${\bf E}$ inside the crystal
\begin{equation}
\nabla \times (\nabla \times {\bf E}) - \omega^2 \widehat{\epsilon} \cdot {\bf E} =0, \label{e21}
\end{equation}
where $\widehat{\epsilon}$  is the dielectric tensor.
Let the principal axis of  $\widehat{\epsilon}$ be ${\bf e}_i$ ($i=1,2,3$) and choose them as
the coordinates axis. Thus
$$
\widehat{\epsilon} = {\rm diag} \{\epsilon_1,\epsilon_2,\epsilon_3 \}
$$
with the convention $\epsilon_1 <\epsilon_2 <\epsilon_3$.

For a plane wave $\propto \exp (i {\bf k}\cdot {\bf x})$:
\begin{equation}
\mathbb{K} \cdot {\bf E}=  0, \label{KE}
\end{equation}
where
\begin{equation}
\mathbb{K} = \widehat{{\bf k}{\bf k}} - {\bf k}^2 \widehat{1} + \omega^2 \widehat{\epsilon}.\label{K}
\end{equation}
Eq \ref{KE} has non-trivial solution if the determinant of the  matrix $\mathbb{K}$ is zero:
\begin{equation}
\Delta \equiv \Big[{\bf k}^2 ~ -\omega^2 (\epsilon_1 + \epsilon_2 + \epsilon_3)\Big] ({\bf k}\cdot {\bf
k'})+\omega^2 \Big[{\bf k'}^2 + \omega^2 \epsilon_1  \epsilon_2  \epsilon_3 \Big]=0,\label{fres}
\end{equation}
which is the Fresnel equation \cite{LL}. Here and in the following we use the convention
$$
{\bf k'} = \widehat{\epsilon} \cdot {\bf k}~,\quad ~~{\bf k''} = \widehat{\epsilon}~^2 \cdot {\bf k}~.
$$

To the vector ${\bf k}$ is associated another vector ${\bf s}$ such that ${\bf k }\cdot {\bf s} =\omega$. Explicitly,
\begin{equation}
{\bf s} = N^{-1}({\bf k }) \Big \{ ({\bf k}\cdot {\bf k'}) ~{\bf k } + \Big[{\bf k }^2 - \omega^2 (\epsilon_1 +
\epsilon_2 + \epsilon_3)\Big] {\bf k' } + \omega^2 {\bf k''} \Big \}~, \label{s}
\end{equation}
where
$$
N ({\bf k })= ({\bf k}\cdot {\bf k'}) ~{\bf k }^2 - \omega^4 \epsilon_1  \epsilon_2  \epsilon_3~.
$$

The optical axis is given by ${\bf k}= \omega {\bf n}$, where
\begin{equation}
n_1 = \sqrt{\frac{\epsilon_3 (\epsilon_2 - \epsilon_1)}{\epsilon_3 - \epsilon_1}}, ~\quad n_2= 0, \quad n_3 =
\sqrt{\frac{\epsilon_1 (\epsilon_3 - \epsilon_2)}{\epsilon_3 -\epsilon_1}} ~,
\end{equation}
and thus
\begin{equation}
{\bf n}^2 = \epsilon_2, \quad {\bf n} \cdot {\bf n'} = \epsilon_1 \epsilon_3, \quad {\bf n'}^2 = \epsilon_1
\epsilon_3 (\epsilon_1- \epsilon_2 + \epsilon_3).\label{nn}
\end{equation}
It can be seen with some simple algebra that the Fresnel determinant can also be written in the form
\begin{equation}
\Delta = ({\bf k}^2 - \omega^2 \epsilon_2)({\bf k} \cdot {\bf k'} - \omega^2 \epsilon_1 \epsilon_3) - \omega^2
(\epsilon_2 - \epsilon_1)(\epsilon_3 - \epsilon_2) k_2^2.
\end{equation}

At the optical axis, both the function $N ({\bf k })$ and the vector term in curly brackets in Eq. (\ref{s}) are
zero and this equation is undefined. This corresponds to the case of internal conical refraction. In this case,
the vector ${\bf s}$ must be  calculated setting ${\bf k} = \omega {\bf n} + \delta {\bf k}$ in the Fresnel
equation (\ref{fres}) and then taking the limit $\delta {\bf k} \rightarrow 0$. Explicitly, setting $\Delta ({\bf
k})\equiv {\rm Det} ~\mathbb{K}$, we find:
$$
\Delta (\omega {\bf n} + \delta {\bf k}) = 2 \omega^3 \Big\{   ({\bf n} \cdot {\bf n'}) {\bf n} 
+ \Big[{\bf n}^2-(\epsilon_1+ \epsilon_2 + \epsilon_3) \Big] {\bf n'} 
$$
$$
+ {\bf n''} \Big\} \cdot \delta {\bf k}
+\omega^2 \delta {\bf k} \cdot \Big\{ ({\bf n} \cdot {\bf n'}) \widehat{1} + 2 (\widehat{{\bf n}  {\bf n'}}
+\widehat{{\bf n'}  {\bf n}}) 
$$
\begin{equation}
+ \Big[{\bf n}^2  - (\epsilon_1+ \epsilon_2 + \epsilon_3) \Big] \widehat{\epsilon} +
\widehat{\epsilon}~^2 \Big\} \cdot \delta {\bf k}
+ \omega O( \delta {\bf k}^3)=0.\label{D}
\end{equation}
This equation can be rewritten in the form
\begin{equation}
2 N (\omega {\bf n} + \delta {\bf k}) ~{\bf s} \cdot \delta {\bf k} + \omega^2 \delta {\bf k} \cdot \mathbb{M}
\cdot \delta {\bf k} =0,
\end{equation}
where
\begin{equation}
N (\omega {\bf n} + \delta {\bf k}) = 2 \omega^3 \Big[\epsilon_1 (\epsilon_2 + \epsilon_3) n_1 \delta k_1
+\epsilon_3 (\epsilon_1 + \epsilon_2) n_3 \delta k_3 \Big]
\end{equation}
and the matrix $\mathbb{M}$ is given by
\begin{equation}
\mathbb{M} = 2 (\widehat{{\bf n} {\bf n'}} + \widehat{{\bf n'} {\bf n}})- (\epsilon_2 - \epsilon_1)(\epsilon_3
-\epsilon_2) ~ \widehat{{\bf e}_2 {\bf e}_2}.
\end{equation}

It then follows that
\begin{equation}
{\bf s} = {\rm lim}_{|\delta {\bf k}|\rightarrow 0} ~\frac{\mathbb{M}  \cdot \delta {\bf k}}{ \delta  {\bf k}
\cdot \mathbb{M} \cdot \delta {\bf k}}
\end{equation}
and clearly
\begin{equation}
{\bf s} \cdot \delta  {\bf k} = 0 = \delta  {\bf k} \cdot \mathbb{M}  \cdot \delta  {\bf k}
\end{equation}

Therefore
\begin{equation}
{\bf s} =\frac{ 2 [({\bf n'} \cdot \delta {\bf k}) {\bf n} + ({\bf n} \cdot \delta {\bf k}) {\bf n'} ]
-(\epsilon_2 - \epsilon_1 )(\epsilon_3 - \epsilon_2 ) ~\delta k_2 }{2[\epsilon_2 ({\bf n'} \cdot \delta {\bf k}) +
\epsilon_1 \epsilon_3 ({\bf n} \cdot \delta {\bf k})]},\label{scone}
\end{equation}
with the conditions
\begin{equation}
4 ({\bf n} \cdot \delta {\bf k}) ~({\bf n'} \cdot \delta {\bf k}) = (\epsilon_2 - \epsilon_1)(\epsilon_3 -
\epsilon_2) (\delta k_2)^2~.\label{scone2}
\end{equation}
Notice that the absolute magnitude of $\delta {\bf k}$ does not appear in these last expressions defining the
vector ${\bf s}$.

\subsection{Geometry of refraction cone}
\begin{figure*}[htbp!]
\begin{center}
\includegraphics[width=0.4\textwidth]{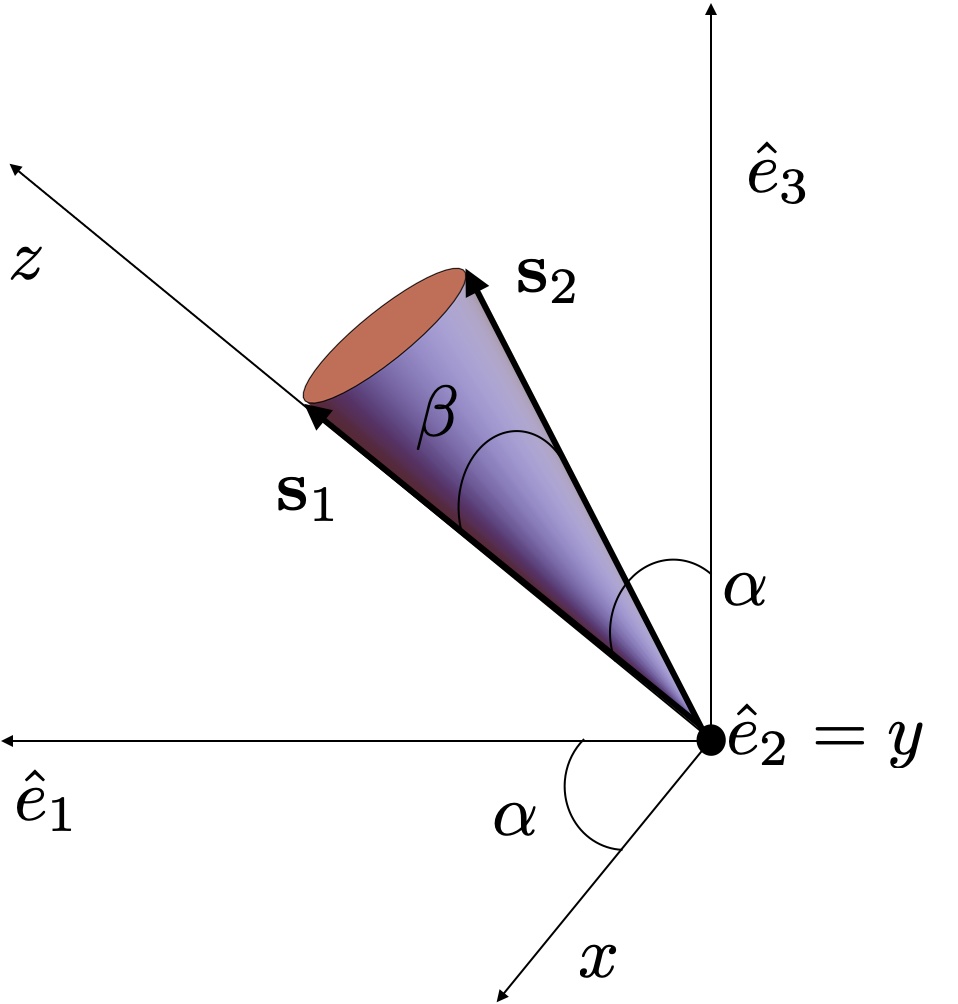}
\end{center}
\caption{ {\bf Schematic of the  refraction cone}. }
\label{Fig1}
\end{figure*}

Eqs. \eqref{scone} and \eqref{scone2} define the vector ${\bf s}$ that sweeps the internal refraction cone. It follows from these equations that the intersections of the cone with the
plane $(3,1)$ is given by two vectors that bound it:
\begin{equation}
{\bf s_1} = \frac{1}{\epsilon_2} {\bf n} \quad {\rm and} \quad {\bf s_2} = \frac{1}{\epsilon_1 \epsilon_3} {\bf
n'},
\end{equation}
corresponding to $\delta k_2 =0$.

Since
$$
{\bf s_1}^2 = \frac{1}{\epsilon_2} = {\bf s_1} \cdot {\bf s_2},
$$
\begin{equation} {\bf s_2}^2 = \frac{1}{\epsilon_3 } + \frac{1}{\epsilon_1} - \frac{\epsilon_2}{\epsilon_1
\epsilon_3},
\end{equation}
it follows that $({\bf s_2} - {\bf s_1}) \cdot {\bf s_1} = 0$.

Consider a cone as depicted in Fig. 1. Choose the $z$ axis in the ${\bf s}$ direction and the $y$ axis in the
${\bf e_2}$ direction. Let $\beta$ be the angle between ${\bf s_1}$ and ${\bf s_2}$, that is
\begin{equation}
\tan \beta =\sqrt{\frac{(\epsilon_2 - \epsilon_1) (\epsilon_3 - \epsilon_2)}{\epsilon_1  \epsilon_3}}.
\end{equation}
The cone is given by the equation
\begin{equation}
f({\bf r}) \equiv x^2 + \tan \beta ~xz + y^2 =0,
\end{equation}
and the vector normal to the cone is
\begin{equation}
{\bf p} \equiv \{x+\frac{1}{2} \tan \beta ~z,~ y,~ \frac{1}{2} \tan \beta ~x \}.\label{d}
\end{equation}
It satisfies the condition
$$
{\bf p} \cdot {\bf s} =0
$$
everywhere on the cone, in accordance with Eq. \ref{scone}.

In this new system of coordinates
\begin{equation}
{\bf n} = \{0,~ 0,~ n \}.
\end{equation}
and
\begin{equation}
{\bf n'} = \{-n' \sin \beta,~ 0,~ n' \cos \beta \},
\end{equation}
where $n$ and $n'$ are given by Eq. \ref{nn}.  It then follows that
\begin{equation}
4 ({\bf n} \cdot {\bf p})({\bf n'} \cdot {\bf p}) = n n' \sin \beta \tan \beta ~p_y^2,\label{nd}
\end{equation}
in accordance with Eq. \ref{scone2}. This last equation implies
\begin{equation}
p_z = \frac{1}{2} \tan \beta \Big( p_x \pm \sqrt{p_x^2 + p_y^2}~\Big).\label{dz=}
\end{equation}

Notice also that the angle $\alpha$ between ${\bf n}$ and the ${\bf e_3}$ axis is given by
\begin{equation}
\tan \alpha =\sqrt{ \frac{\epsilon_3 (\epsilon_2 - \epsilon_1)}{\epsilon_1 (\epsilon_3 - \epsilon_2)}}.
\end{equation}
Accordingly, the matrix $\widehat{\epsilon}$ in this system of coordinates is 
\begin{equation}
\widehat{\epsilon} =
\begin{pmatrix}
  \epsilon_1 \cos^2 \alpha + \epsilon_3 \sin^2 \alpha & 0 & -(\epsilon_3 - \epsilon_1) \sin \alpha \cos \alpha  \\
 0 & \epsilon_2 & 0 \\
  -(\epsilon_3 - \epsilon_1) \sin \alpha \cos \alpha & 0 & \epsilon_1 \sin^2 \alpha + \epsilon_3 \cos^2 \alpha
\end{pmatrix}.\label{eps}
\end{equation}
The following  relation is useful:
\begin{equation}
\epsilon_1 \sin^2 \alpha + \epsilon_3 \cos^2 \alpha =\epsilon_{zz} = \frac{\epsilon_1 \epsilon_3}{\epsilon_2}.\label{eee}
\end{equation}
Thus, for any vector ${\bf k}$, the scalar product ${\bf k}^2= k_x^2 +k_y^2+k_z^2$ and we also have
$$
{\bf k}\cdot {\bf k'} = \Big(\epsilon_1+ \epsilon_3 -\frac{\epsilon_1 \epsilon_3}{\epsilon_2}\Big)k_x^2
+\epsilon_2 k_y^2 + \frac{\epsilon_1 \epsilon_3}{\epsilon_2} k_z^2
$$
$$
-\frac{2}{\epsilon_2} ~\sqrt{\epsilon_1 \epsilon_3 (\epsilon_2 -\epsilon_1)(\epsilon_3 -\epsilon_2)}~k_x k_z
$$
$$
= \frac{\epsilon_1 \epsilon_3}{\epsilon_2} \Big(  ~k_z - \tan \beta~k_x \Big)^2 +\epsilon_2 (k_x^2 +k_y^2).
$$

Accordingly, the Fresnel equation implies
\begin{eqnarray}
\Big[k_z^2 + k_{\bot}^2- \epsilon_2 \omega^2\Big] \Big[(k_z -\tan \beta ~k_x)^2 - \epsilon_2 (\omega^2 -
\frac{\epsilon_2}{\epsilon_1 \epsilon_3} k_{\bot}^2)\Big]
\nonumber\\
- \epsilon_2 \omega^2 \tan^2 \beta ~k_y^2 =0.
 \label{F2}
\end{eqnarray}
As a polynomial in $k_z$, it has four (real) roots: two positive and two negative ones.

Notice that for $\beta \ll 1$, the above equation has the following four solutions for $k_z$:
$$
 K_-= \pm \sqrt{\epsilon_2 \omega^2 -k_{\bot}^2} +O(\beta^2)
$$
\begin{equation}
 K_+= \pm \sqrt{\epsilon_2 \omega^2- \frac{\epsilon_2^2}{\epsilon_1 \epsilon_3}  k_{\bot}^2} +~\beta ~k_x +O(\beta^2).\label{betita}
\end{equation}

In the particular case of a uniaxial crystal such as, say,   $\epsilon_1 = \epsilon_2 \equiv \epsilon$, we have
$\beta =0$ and the four roots are given by
$$
k_z^2 = \epsilon ~\omega^2 - k^2_{\bot}
$$
and
$$
k_z^2 = \epsilon ( \omega^2 - \epsilon^{-1}_3 k_{\bot}^2).
$$
The first root corresponds to the ordinary wave and the second to the extraordinary wave.

In the following, we take $K_+$ and $K_-$ as the two positive roots of \eqref{F2}, corresponding to propagation in
the positive $z$ direction inside the crystal.

\subsection{Fourier transform}

The general solution of Eq. (\ref{2.1}) can be written in the form (the term $e^{i\omega t}$ is not included for
simplicity)
\begin{equation}
{\bf E} ({\bf r})= \frac{1}{(2\pi)^{3/2}}  \int e^{i {\bf k} \cdot {\bf r}}  \delta (\Delta) \widetilde{{\bf
F}}({\bf k}) ~d{\bf k}  ,\label{fou}
\end{equation}
where $\Delta$ is the Fresnel determinant and $\widetilde{{\bf F}}({\bf k})$ are functions to be determined by
boundary conditions, as shown in the following.

Accordingly the Fourier transform \eqref{fou} reduces to a two-dimensional integral:
$$
{\bf E} ({\bf r})= \frac{1}{2\pi}  \int \int  e^{i k_x x +i k_y y} 
$$
\begin{equation}
\times\Big[ e^{i K_+ z }~ \widetilde{{\bf E}}^+(k_x,
k_y)+e^{i K_- z } ~\widetilde{{\bf E}}^-(k_x, k_y)\Big]~dk_x ~ dk_y,  \label{fou2}
\end{equation}
where $\widetilde{{\bf E}}^{\pm}(k_x, k_y)$ are to be determined by the boundary conditions. A similar equation
applies to ${\bf D}$ with $\widetilde{{\bf D}}^{\pm} = \hat{\epsilon} \cdot  \widetilde{{\bf E}}^{\pm}$. As for
the magnetic field, it is
$$
{\bf B} ({\bf r})= ~\frac{1}{2\pi  \omega } \int \int  e^{i (k_x x +k_y y)} \Big[e^{i K_+ z }( {\bf k}_{\bot} +
K_+ {\bf e_z})\times \widetilde{{\bf E}}^+
$$
\begin{equation}
+e^{i K_- z }( {\bf k}_{\bot} + K_- {\bf e_z})\times \widetilde{{\bf E}}^-  \Big] ~dk_x dk_y.
\end{equation}

\section{Reflection and refraction}

In order to study the reflection and refraction of the waves, we write
the electric vector ${\bf E}$ in vacuum (that is, for $z<0$) in the
form
$$
{\bf E}(x,y,z) = \frac{1}{2\pi} \int  dk_x dk_y~ e^{ ik_x x +ik_y y}
$$
\begin{equation}
\Big[e^{i k_z z} ~\widetilde{{\bf E}}^I (k_x,k_y) + e^{-i k_z z}
~\widetilde{{\bf E}}^R (k_x,k_y) \Big]~,\label{EVIR}
\end{equation}
where $k_z= (\omega^2 -k_x^2 -k_y^2)^{1/2}$ and $\widetilde{{\bf E}}^{(I,R)} (k_x,k_y)$ are the two-dimensional
Fourier transforms of the electric field components of the incident
and reflected waves, ${\bf E}^{(I,R)}(x,y,0^-)$ at the interface;
similar equations apply to the magnetic field component.

The boundary conditions imply the continuity of $E_x$, $E_y$, $B_x$
and $B_y$ at the interface $z=0$ (the continuity conditions on $D_z$
and $B_z$ are not independent since, from the Maxwell equations, $i
\omega D_z = \partial_y B_x - \partial_x B_y$ and $i \omega B_z =
\partial_x E_y - \partial_y E_x$). It is convenient to express each
Fourier transformed component of ${\bf B}$ and $E_z$ in the vacuum
region in terms of only $E_x$ and $E_y$ using the Maxwell equations.
For the incident field (see \cite{hj2}):
\begin{eqnarray}
&\widetilde{E}^I_z &= - ~\frac{1}{k_z} \Big(k_x \widetilde{E}^I_x +
k_y
\widetilde{E}^I_y \Big) \\
&\widetilde{B}^I_x &= - ~\frac{1}{k_z\omega} \Big[k_x k_y
\widetilde{E}^I_x + (k_y^2 + k_z^2)
\widetilde{E}^I_y \Big] \\
&\widetilde{B}^I_y &=  \frac{1}{k_z\omega} \Big[(k_x ^2 + k_z^2)
\widetilde{E}^I_x + k_x k_y \widetilde{E}^I_y \Big]\\
&\widetilde{B}^I_z &=  ~\frac{1}{\omega} \Big(-k_y \widetilde{E}^I_x
+ k_x \widetilde{E}^I_y \Big).
\end{eqnarray}
These equations can be rewritten in terms of a $2 \times 2$ dyad as
\begin{equation}
{\bf e}_z \times \widetilde{{\bf B}}^I = -k_z \omega ~(\omega^2
\widehat{1} - \widehat{{\bf k}_{\bot} {\bf k}_{\bot}})^{-1}~ \widetilde{{\bf
E}}^I_{\bot}
\end{equation}
Here and in the following, ${\bf V}_{\bot}=(V_x,V_y)$ for any vector
${\bf V}$ and also
$$
\widehat{{\bf k}_{\bot} {\bf k}_{\bot}} \equiv \left(
                                            \begin{array}{cc}
                                              k_x^2 & k_xk_y \\
                                              k_xk_y & k_y^2 \\
                                            \end{array}
                                          \right).
$$

For the reflected field, it is only necessary to change the sign of
$k_z$. Accordingly
\begin{equation}
{\bf e}_z \times (\widetilde{{\bf B}}^I + \widetilde{{\bf B}}^R) =
-k_z \omega ~(\omega^2 \widehat{1} - \widehat{{\bf k}_{\bot} {\bf k}_{\bot}})^{-1}~ (\widetilde{{\bf E}}^I_{\bot} - \widetilde{{\bf
E}}^R_{\bot}) ~,
\end{equation}
and the boundary conditions take the form
\begin{equation}
\widetilde{{\bf E}}^I_{\bot} + \widetilde{{\bf E}}^R_{\bot}
=\widetilde{{\bf E}}^+_{\bot} + \widetilde{{\bf E}}^-_{\bot}~,\label{e42}
\end{equation}
and
\begin{equation}
\widetilde{{\bf E}}^I_{\bot} - \widetilde{{\bf E}}^R_{\bot} =
-~\frac{1}{k_z \omega}~ (\omega^2 \widehat{1} - \widehat{{\bf k}_{\bot} {\bf k}_{\bot}}) ~[{\bf e}_z \times (\widetilde{{\bf B}}^+ +
\widetilde{{\bf B}}^-)]~.\label{e43}
\end{equation}

At this point, it is convenient to define
$$
{\bf F} \equiv \widetilde{{\bf E}}^+ + \widetilde{{\bf E}}^-
$$
$$
k_z {\bf G} \equiv K_+\widetilde{{\bf E}}^+ + K_-\widetilde{{\bf E}}^- ,
$$
and accordingly,
$$
\widetilde{{\bf E}}^+ = \frac{1}{K_--K_+}(K_- {\bf F} -k_z{\bf G})
$$
\begin{equation}
\widetilde{{\bf E}}^- = \frac{1}{K_--K_+}(-K_+ {\bf F} +k_z {\bf G});\label{EEFG}
\end{equation}
also
\begin{equation}
K_+^2\widetilde{{\bf E}}^+ + K_-^2\widetilde{{\bf E}}^- = -K_+ K_- {\bf F} + (K_+ + K_-)k_z {\bf G}. \label{FG}
\end{equation}

Since $i\omega {\bf B} =\nabla \times {\bf E}$, equation  \eqref{e43} takes the explicit form
\begin{equation}
\widetilde{{\bf E}}^I_{\bot} - \widetilde{{\bf E}}^R_{\bot} = -~\frac{1}{ \omega^2}\Big[ k_z F_z{\bf k}_{\bot}
-(\omega^2 \widehat{1} - \widehat{{\bf k}_{\bot} {\bf k}_{\bot}}) {\bf G}_{\bot}\Big]
\label{e3}.
\end{equation}

From this last equation and \eqref{e42} and \eqref{e43}, we eliminate ${\bf E}^R_{\bot}  $ and get
\begin{equation}
2\widetilde{{\bf E}}^I_{\bot}  = {\bf F}_{\bot} -~\frac{1}{ \omega^2}\Big[ k_z F_z{\bf k}_{\bot}
-(\omega^2 \widehat{1} - \widehat{{\bf k}_{\bot} {\bf k}_{\bot}})  {\bf G}_{\bot}\Big]
\label{d}.
\end{equation}

This equation must be supplemented with \eqref{e21},  which now takes the form
$$
({\bf k}_{\bot} \cdot {\bf F}_{\bot} +k_z G_z) {\bf k}_{\bot} + (K_+ K_- - k_{\bot}^2) {\bf F}_{\bot} 
$$
\begin{equation}
- (K_+ +K_-) k_z {\bf G}_{\bot} +\omega^2 ( \hat{\epsilon} \cdot {\bf F})_{\bot} =0 \label{a}
\end{equation}
\begin{equation}
k_z {\bf k}_{\bot} \cdot {\bf G}_{\bot} - k_{\bot}^2 F_z  +\omega^2 ( \hat{\epsilon} \cdot {\bf F})_z =0 \label{b}.
\end{equation}

We also have the condition $\nabla \cdot {\bf D} =0$ which implies
\begin{equation}
 {\bf k}_{\bot} \cdot ( \hat{\epsilon} \cdot {\bf F}) + k_z   ( \hat{\epsilon} \cdot {\bf G})_z =0 \label{c}.
\end{equation}

Thus we have a set of six equations for the six components of ${\bf F}$ and ${\bf G}$.

It is convenient to rewrite \eqref{a} and \eqref{b} in the form
\begin{equation}
\mathbb{A} \left(
             \begin{array}{c}
               F_x \\
               F_y \\
               F_z \\
             \end{array}
           \right)  +
           \mathbb{B}k_z \left(
             \begin{array}{c}
               G_x \\
               G_y \\
               G_z \\
             \end{array}
           \right)=0,
\end{equation}
and \eqref{d} and \eqref{c} as
\begin{equation}
\mathbb{C} \left(
             \begin{array}{c}
               F_x \\
               F_y \\
               F_z \\
             \end{array}
           \right)  +
           \mathbb{D}k_z \left(
             \begin{array}{c}
               G_x \\
               G_y \\
               G_z \\
             \end{array}
           \right)=2 \omega^2
           \left(
             \begin{array}{c}
               \widetilde{E}^I_x \\
               \widetilde{E}^I_y \\
               0 \\
             \end{array}
           \right),
\end{equation}
where $\mathbb{A}$, $\mathbb{B}$, $\mathbb{C}$, and $\mathbb{D}$ are $3 \times 3$ matrices. Explicitly,
\begin{equation}
\mathbb{A}=\omega^2 \hat{\epsilon} -k_{\bot}^2 \hat{1}+ \left(
              \begin{array}{ccc}
                K_+K_- +k_x^2 & k_x k_y & 0 \\
                k_x k_y & K_+K_- +k_y^2 & 0 \\
                0 & 0 & 0 \\
              \end{array}
            \right),
            \end{equation}
where $\hat{\epsilon}$ is given by \eqref{eps}   and $\hat{1}$ is the $3 \times 3$ unit matrix,
\begin{equation}
\mathbb{B}=            \left(
                                  \begin{array}{ccc}
                                    -(K_++K_-) & 0 & k_x \\
                                    0 & -(K_++K_-) & k_y \\
                                    k_x & k_y & 0 \\
                                  \end{array}
                                \right),
\end{equation}
\begin{equation}
\mathbb{C}=            \left(
                                  \begin{array}{ccc}
                                    \omega^2 & 0 & -k_x k_z\\
                                    0 & \omega^2 & -k_y k_z\\
                                    k_x \epsilon_{xx} & k_y \epsilon_{yy} & k_x \epsilon_{xz} \\
                                  \end{array}
                                \right),
\end{equation}
\begin{equation}
\mathbb{D}=            \left(
                                  \begin{array}{ccc}
                                    \omega^2-k_x^2 & -k_x k_y & 0\\
                                    -k_x k_y & \omega^2-k_y^2 & 0\\
                                    k_z \epsilon_{zx} & 0 & k_z \epsilon_{zz} \\
                                  \end{array}
                                \right).
\end{equation}

Summing up, we can obtain $\widetilde{{\bf E}}^+$ and $\widetilde{{\bf E}}^-$  from the following set of equations:
\begin{equation}
\Big(\mathbb{A} + K_+~ \mathbb{B}\Big) \left(
             \begin{array}{c}
               \widetilde{E}_x^+ \\
               \widetilde{E}_y^+ \\
               \widetilde{E}_z^+ \\
             \end{array}
           \right)  +\Big(\mathbb{A} + K_- ~\mathbb{B}\Big) \left(
             \begin{array}{c}
               \widetilde{E}_x^- \\
               \widetilde{E}_y^- \\
               \widetilde{E}_z^- \\
             \end{array}
           \right) =0,\label{111}
\end{equation}
\begin{equation}
\Big(\mathbb{C} + K_+~ \mathbb{D}\Big) \left(
             \begin{array}{c}
               \widetilde{E}_x^+ \\
               \widetilde{E}_y^+ \\
               \widetilde{E}_z^+ \\
             \end{array}
           \right)  +\Big(\mathbb{C} + K_- ~\mathbb{D}\Big) \left(
             \begin{array}{c}
               \widetilde{E}_x^- \\
               \widetilde{E}_y^- \\
               \widetilde{E}_z^- \\
             \end{array}
           \right)=2 \omega^2
           \left(
             \begin{array}{c}
               \widetilde{E}^I_x \\
               \widetilde{E}^I_y \\
               0 \\
             \end{array}
           \right).\label{222}
\end{equation}

It is worth noticing that in the particular case $K^+ =K_- \equiv K$, which may occur for $k_{\bot} \sim \beta$, the determinant of $ \mathbb{A}+ K\mathbb{B}$ is zero, and therefore equation \eqref{111} is undetermined; however, \eqref{222} yields the solution for $\widetilde{{\bf E}}^++\widetilde{{\bf E}}^-$, which is the combination appearing in the Fourier transform \eqref{fou2} if $K^+ =K_-$. In any case, we do not have this problem in the particular examples considered hereafter.

\section{Numerical evaluations}

In this section, we present the numerical evaluations. For definiteness, we choose the parameters of a KTP crystal and perform the integrations for two Gaussian beams, linearly and circularly polarized, and a zero-order Bessel beams. The results are shown in figures 2,3, and 4, where the unit of length is taken as $k^{-1}=  \lambda /2\pi =1$.

\subsection{KTP crystal}

For a KTP crystal, such as the one used in Ref.  \cite{nature},
$$
\epsilon_1= 3.1609, \quad \epsilon_2=3.1994, \quad \epsilon_3=3.5672,
$$
and therefore $\tan \beta =0.0354$ (also $\epsilon_2^2 /(\epsilon_1 \epsilon_3) = 0.9078~ {\rm and}~ \epsilon_2 /(\epsilon_1 \epsilon_3) = 0.2837$)

\begin{figure*}[htbp!]
\begin{center}
\includegraphics[width=0.4\textwidth]{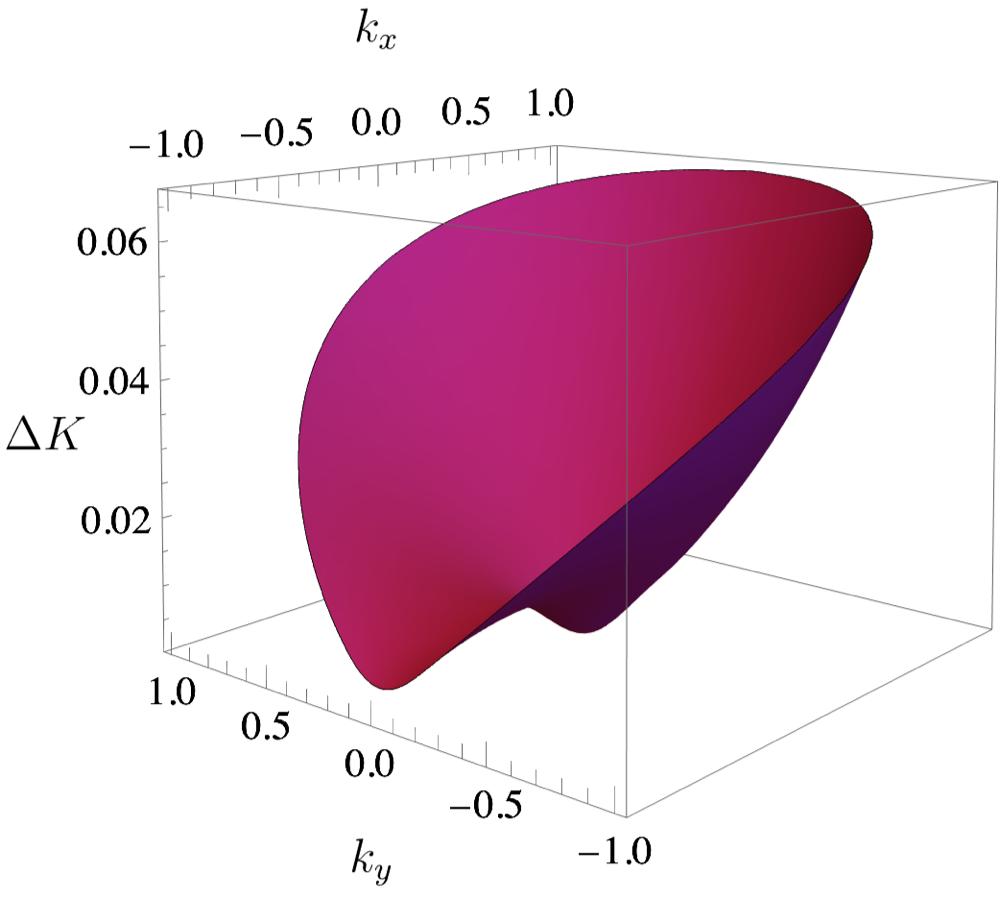}
\end{center}
\caption{ {\bf Difference $\Delta K=K_+-K_-$}. The difference in the exact numerical solutions $K_\pm$ of \eqref{F2}. $\Delta K\neq 0$ for all values of $0\leq k_\perp\leq 1$. Parameters used are for the biaxial  KTP crystal considered in the main text.  }
\label{Fig2}
\end{figure*}

\subsection{Gaussian beam}

Consider a Gaussian beam polarized in the $x$ direction and moving along the $z$ axis. It has the form
\begin{equation}
{\bf E}= E_0 G({\bf r})\hat{{\bf e}}_x,
\end{equation}
with
\begin{equation}
G({\bf r})= \frac{1}{q(z)}
e^{i\omega z-i \omega r^2/2q(z) },
\end{equation}
where $E_0$ is the amplitude, $q(z)=z+iz_R$ and $z_R$ is the Rayleigh range, defined as
$$z_R= \omega w_0^2 /2~,$$
in terms of the waist radius $w_0$. 

As a Fourier transform, we have
$$
G({\bf r})= \frac{1}{2\pi i \omega} \int \int \int d{\bf k}~ e^{i{\bf k} \cdot {\bf r} - z_R k_{\bot}^2 /2\omega} ~\delta(k_z -\omega -k_{\bot}^2/2\omega)
$$
\begin{equation}
= \frac{1}{2\pi i \omega} \int \int dk_x~dk_y ~ e^{ i k_x x+ik_y y +i\omega z + i k_{\bot}^2 q(z)/2\omega } .\label{fg}
\end{equation}

It then follows that
\begin{equation}
\tilde{E}^I_x= \frac{E_0 }{ i \omega} ~ e^{- z_R  ~k_{\bot}^2 /2\omega }, \quad \tilde{E}^I_y= 0 .\label{fgauss}
\end{equation}
For simplicity, the waist of the beam is assumed to coincide with the surface of the crystal; thus, we set $z=0$ in \eqref{fg}.

The above values  must be substituted in Eqs. \eqref{111} and \eqref{222}, and then the field inside the crystal can be calculated with \eqref{fou2}.
Explicitly, this integral is in polar coordinates ${\bf r}=(r,\phi,z)$
$$
{\bf E} (r,\phi,z)= \frac{1}{2\pi}  \int_0^{2\pi} d\phi' \int_0^{\omega} dk_{\bot} k_{\bot} e^{i k_{\bot}r\cos (\phi-\phi')}
$$
\begin{equation}
\times \Big[ e^{i K_+ z }~ \widetilde{{\bf E}}^+(k_{\bot} ,\phi')+e^{i K_- z } ~\widetilde{{\bf E}}^-(k_{\bot} ,\phi')\Big]  ,\label{g2}
\end{equation}
with
$$
k_z =\sqrt{\omega^2 -k_{\bot}^2}
$$
in all the formulas.

We can now use  the equations in the previous section, with $k_x=k_{\bot}\cos \phi'$, $k_y=k_{\bot}\sin \phi'$, and $\widetilde{{\bf E}}^I$ given by \eqref{fgauss} and $\widetilde{E}^I_y=0$.

Another possibility is a circularly polarized beam:
\begin{equation}
{\bf E}\propto ({\bf \hat{e}_x}-i{\bf \hat{e}_y}).
\end{equation}
The Fourier transform is then as \eqref{g2}, but with
\begin{equation}
 \widetilde{E}^I_x= -i \widetilde{E}^I_y, \label{fgausscir}
\end{equation}
and therefore
$$
(k_x -i k_y) \widetilde{E}^I_x =-k_z \widetilde{E}^I_z.
$$

\subsection{Bessel beam}

For a Bessel beam of order 0, propagating along the $z$ axis, we have \cite{hj} (at $z=0$)
$$
{\bf \tilde{E}^I} ({\bf k})= {\bf \tilde{E}^I}(k_{\bot},\phi, k_z)={\bf \tilde{e}^I} (\phi) ~\delta(k_z-\omega \cos \zeta )~\delta (k_{\bot} - \omega \sin \zeta ),
$$
where
$$
{\bf \tilde{e}^I} ( \phi)= \frac{i}{4 \omega \sin \zeta}  \Big[ ({\cal
E} + i {\cal B} \cos \zeta )  e^{-i\phi} ({\bf \hat{e}_x} +i{\bf \hat{e}_y})
$$
\begin{equation}
-~({\cal E} - i {\cal B} \cos \zeta ) e^{i\phi} ({\bf \hat{e}_x} -i{\bf \hat{e}_y}) -2i{\cal B}~\sin \zeta~ {\bf
\hat{e}_z} \Big]  ,
\end{equation}
and $\zeta$ is the axicon angle.

Accordingly, in the Fresnel equation \eqref{F2} and all the above equations, it is enough to set
$$k_x= \omega \sin \zeta \cos \phi' ,\quad  \quad k_y= \omega \sin \zeta \sin \phi',
$$
and solve the Fourier integral, with $\phi'$ as the only variable.

In cylindrical coordinates,
$$
x=r \cos \phi , \quad y=r\sin \phi,
$$
we have \emph{inside the crystal}, according to \eqref{fou2},
$$
{\bf E^{in}}({\bf r})=\frac{\omega }{2\pi } \int_0^{2\pi} d\phi' ~e^{i \omega  \sin \zeta ~r ~\cos (\phi'-\phi) }
$$
\begin{equation}
~\times\Big[ e^{i K_+ z }~ \widetilde{{\bf E}}^+(\phi')+e^{i K_- z } ~\widetilde{{\bf E}}^-(\phi')\Big].
\end{equation}
In this last integral, it is understood that $k_{\bot} = \omega \sin \zeta$ and $k_z=\omega \cos \zeta$, and
therefore the functions in the integral depend on the integration variable $\phi'$ only, and on the distance $z$
inside the crystal through the exponents.

\subsection{Simulation results}
 In order to see the propagation of the beams, we first solve \eqref{F2} numerically with the parameters of the KTP crystal and obtain the solutions for $K_\pm$. The numerical solutions are computed in the relevant interval of parameters, $0\leq k_\perp\leq 1$. The full numerical solution is needed, since a perturbation treatment of the equations leads to spurious zeros in $\Delta K=K_+-K_-$. However, as seen in figure \ref{Fig2}, this quantity is small but always positive, which guaranties that the simultaneous numerical solutions of the systems \eqref{111} and \eqref{222} are well defined.

\begin{figure*}[htbp!]
\begin{center}
\includegraphics[width=0.5\textwidth]{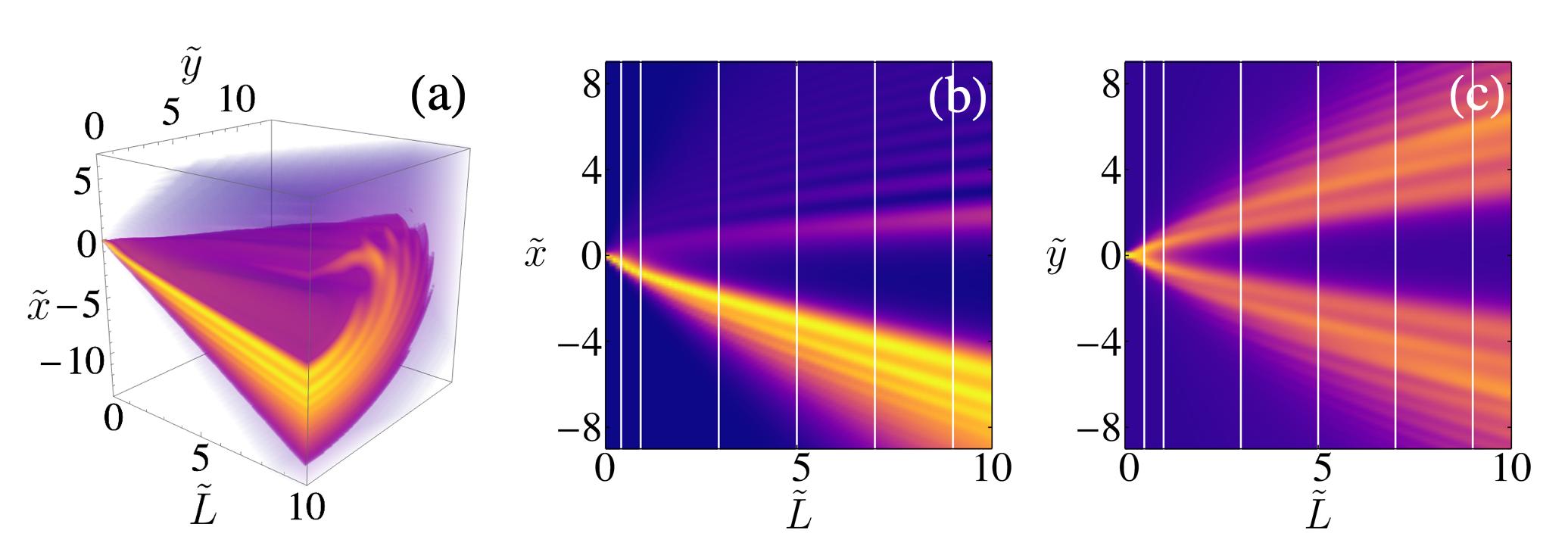}
\end{center}
\caption{ {\bf Propagation of the intensity $|\mathbf{E}^{\mathbf{in}}|^2$ and its projections for an incident gaussian beam polarized in the $x$ axis}. (a) Propagation along the crystal length $\tilde{L}=L\times 10^{-3}$, $\tilde{x}=x\times10^{-2}$ and $\tilde{y}=y\times10^{-2}$. (b) Projection of the propagation for $\tilde{y}=0$. (c) Projection of the propagation for $\tilde{x}=-2.5 \tilde{L}\times 10^{-1}$, the approximate axis for the maxima in the intensity profile. Crystal parameters are the same as in figure \ref{Fig2}. The waist of the gaussian beam is  $w_0=10$.
}
\label{Fig3}
\end{figure*}

\begin{figure*}[htbp!]
\begin{center}
\includegraphics[width=0.5\textwidth]{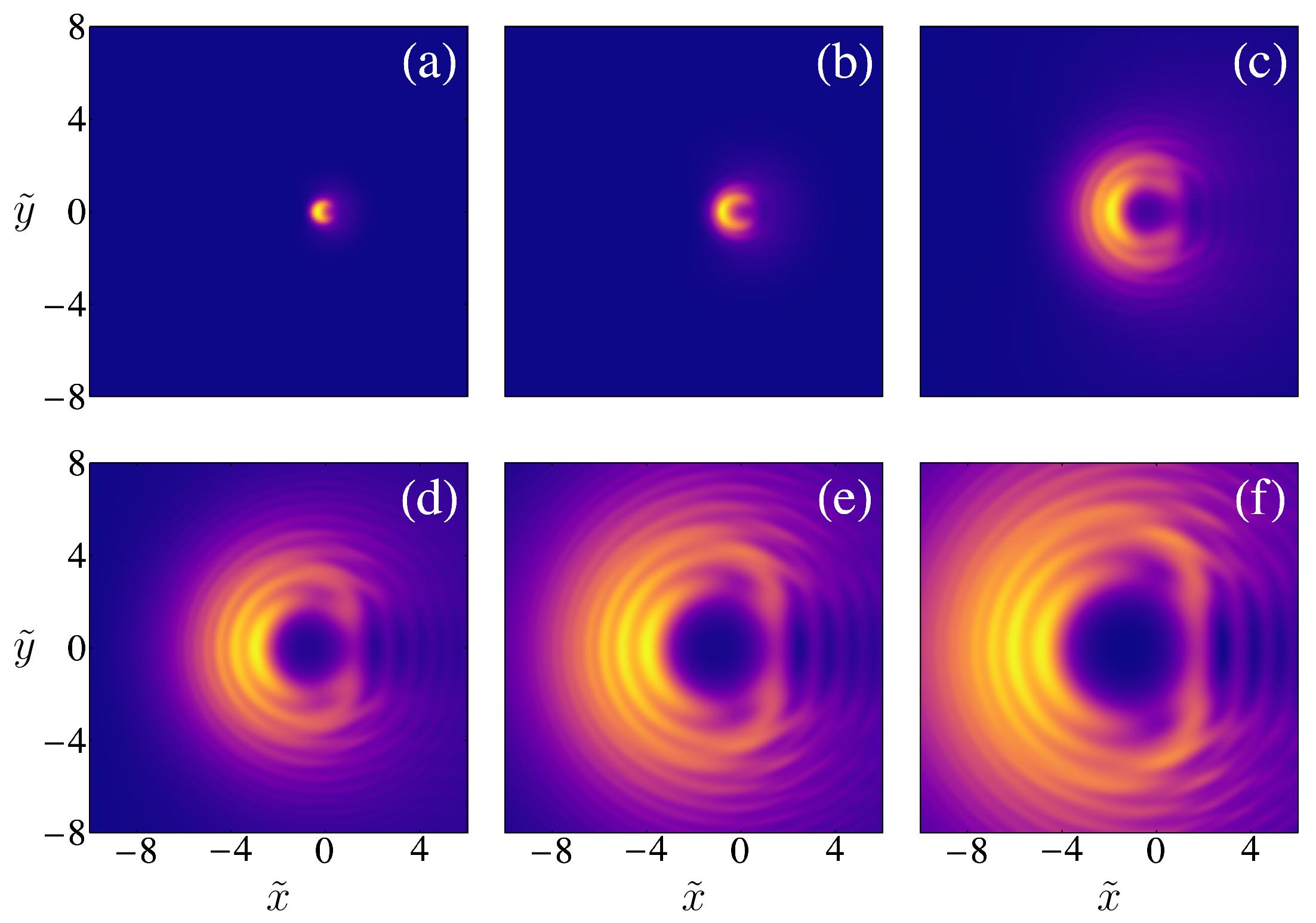}
\end{center}
\caption{ {\bf Transverse planes at different propagation distances $\tilde{L}$ for or an incident gaussian beam polarized in the $x$ axis}. The distance propagated along the crystal corresponds to the white lines in \ref{Fig3} (b) and (c). The distances are $\tilde{L}=$ 0.5(a), 1(b), 3(c), 5(d), 7(e), 9(f). Parameters are the same as in figure \ref{Fig3}.
 }
\label{Fig4}
\end{figure*}

\begin{figure*}[htbp!]
\begin{center}
\includegraphics[width=0.5\textwidth]{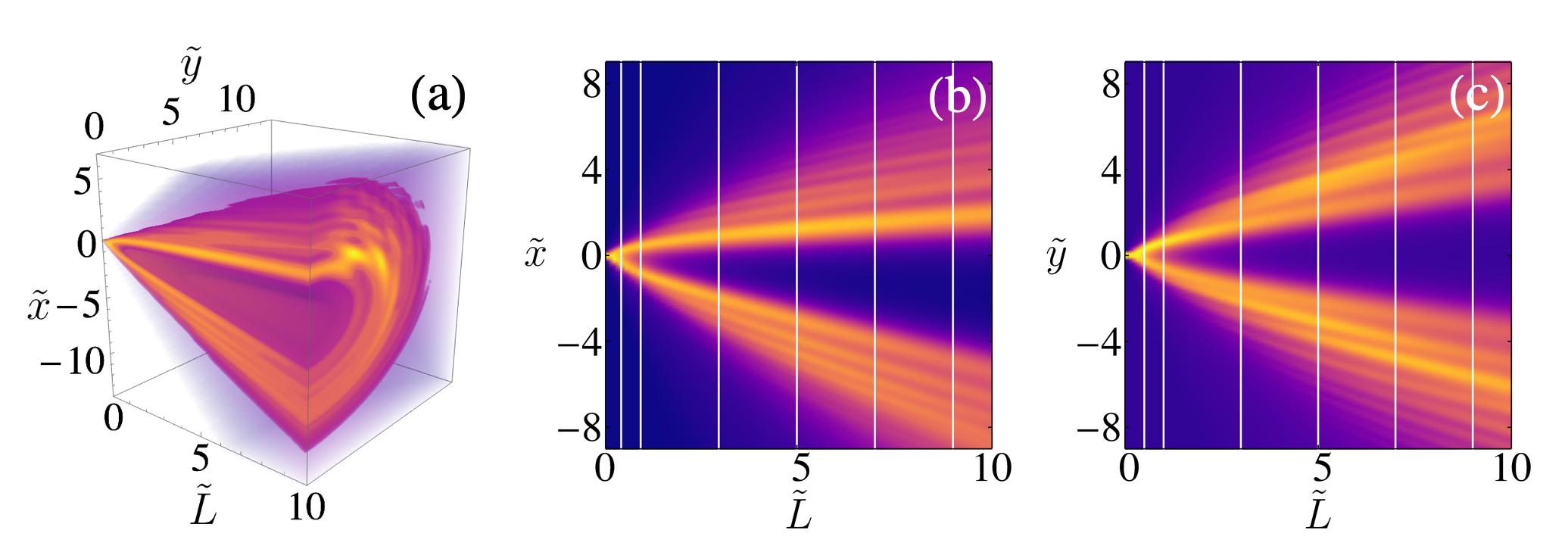}
\end{center}
\caption{ {\bf Propagation of the intensity $|\mathbf{E}^{\mathbf{in}}|^2$ and its projections for an incident  circularly polarized gaussian beam }. (a) Propagation along the crystal length $\tilde{L}=L\times 10^{-3}$, $\tilde{x}=x\times10^{-2}$ and $\tilde{y}=y\times10^{-2}$. (b) Projection of the propagation for $\tilde{y}=0$. (c) Projection of the propagation for $\tilde{x}=-2.5\tilde{L}\times 10^{-1}$, the approximate axis for the maxima in the intensity profile. Crystal parameters are the same as in figure \ref{Fig2}.  The waist of the gaussian beam is  $w_0=10$.
}
\label{Fig5}
\end{figure*}

\begin{figure*}[htbp!]
\begin{center}
\includegraphics[width=0.5\textwidth]{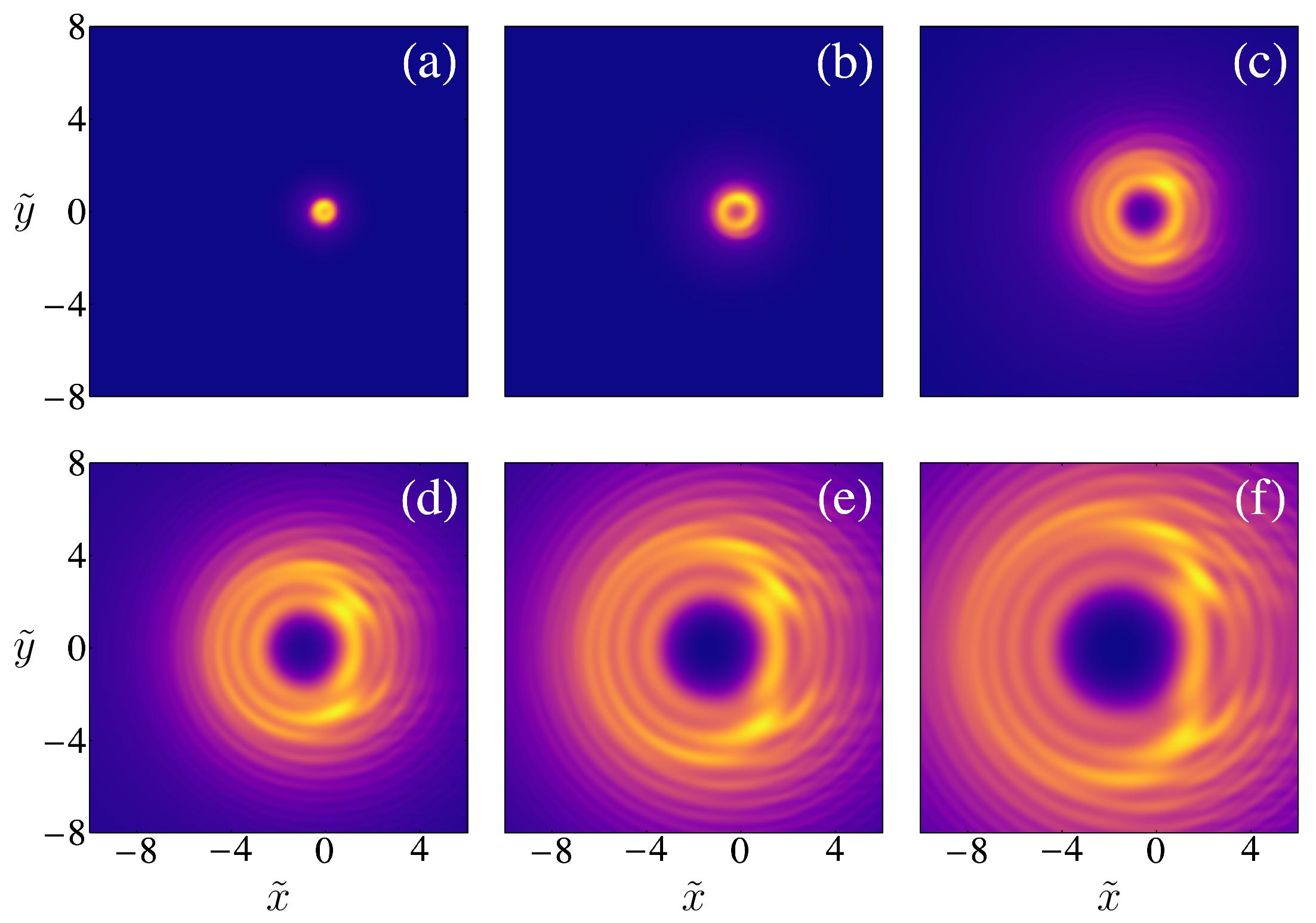}
\end{center}
\caption{ {\bf Transverse planes at different propagation distances $\tilde{L}$ for or an incident gaussian beam circularly polarized}. The distance propagated along the crystal corresponds to the white lines in  Fig. \ref{Fig5} (b) and (c). The distances are $\tilde{L}=$ 0.5(a), 1(b), 3(c), 5(d), 7(e), 9(f). Parameters are the same as in figure \ref{Fig5}.
 }
\label{Fig6}
\end{figure*}

\begin{figure*}[htbp!]
\begin{center}
\includegraphics[width=0.5\textwidth]{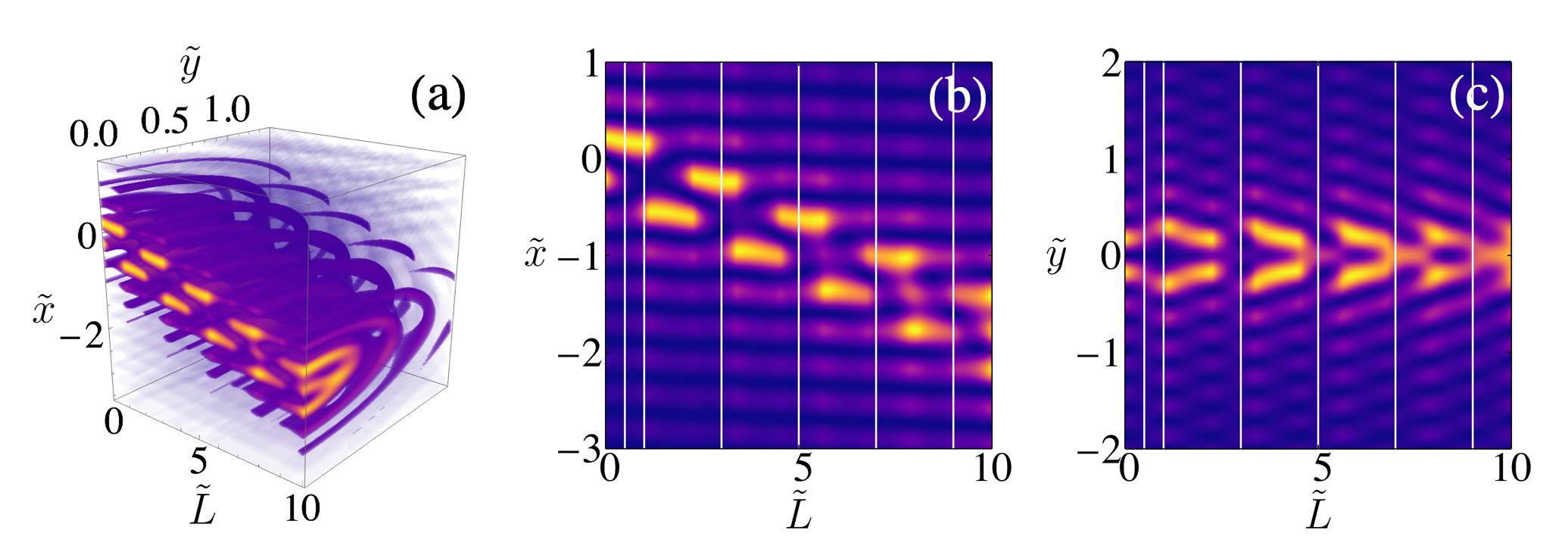}
\end{center}
\caption{ {\bf Propagation of the intensity $|\mathbf{E}^{\mathbf{in}}|^2$ and its projections for an incident linearly polarized Bessel beam }. (a) Propagation along the crystal length $\tilde{L}=L\times 10^{-3}$, $\tilde{x}=x\times10^{-2}$ and $\tilde{y}=y\times10^{-2}$. (b) Projection of the propagation for $\tilde{y}=0$. (c) Projection of the propagation for $\tilde{x}=-1.5 \tilde{L}\times 10^{-1}$, the approximate axis for the maxima in the intensity profile. Crystal parameters are the same as in figure \ref{Fig2}.
}
\label{Fig7}
\end{figure*}

\begin{figure*}[htbp!]
\begin{center}
\includegraphics[width=0.5\textwidth]{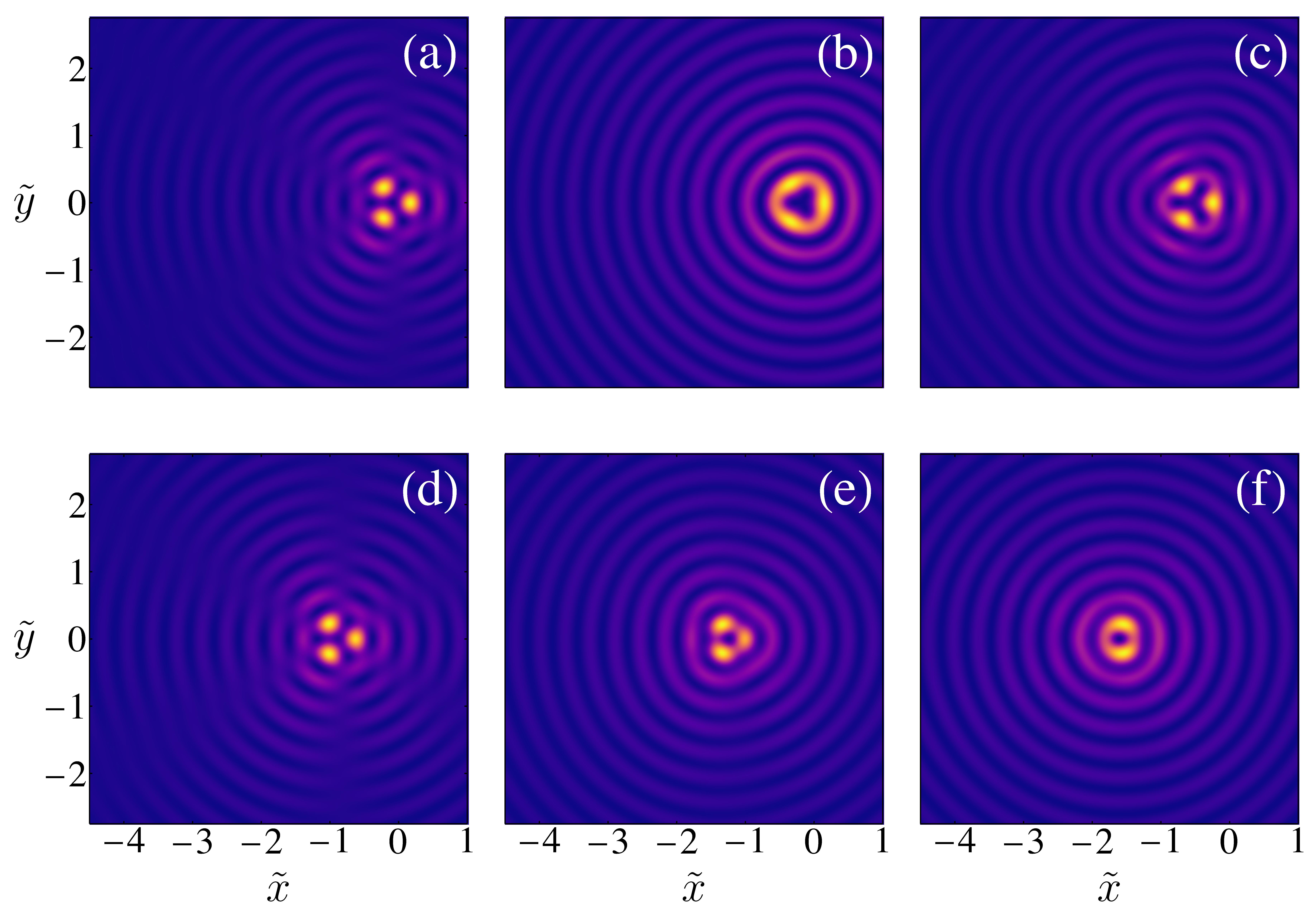}
\end{center}
\caption{ {\bf Transverse planes at different propagation distances $\tilde{L}$ for or an incident linearly polarized Bessel beam}. The distance propagated along the crystal corresponds to the white lines in Fig.  \ref{Fig7} (b) and (c). The distances are $\tilde{L}=$ 0.5(a), 1(b), 3(c), 5(d), 7(e), 9(f). Parameters are the same as in figure \ref{Fig7}.
 }
\label{Fig8}
\end{figure*}

\begin{figure*}[htbp!]
\begin{center}
\includegraphics[width=0.5\textwidth]{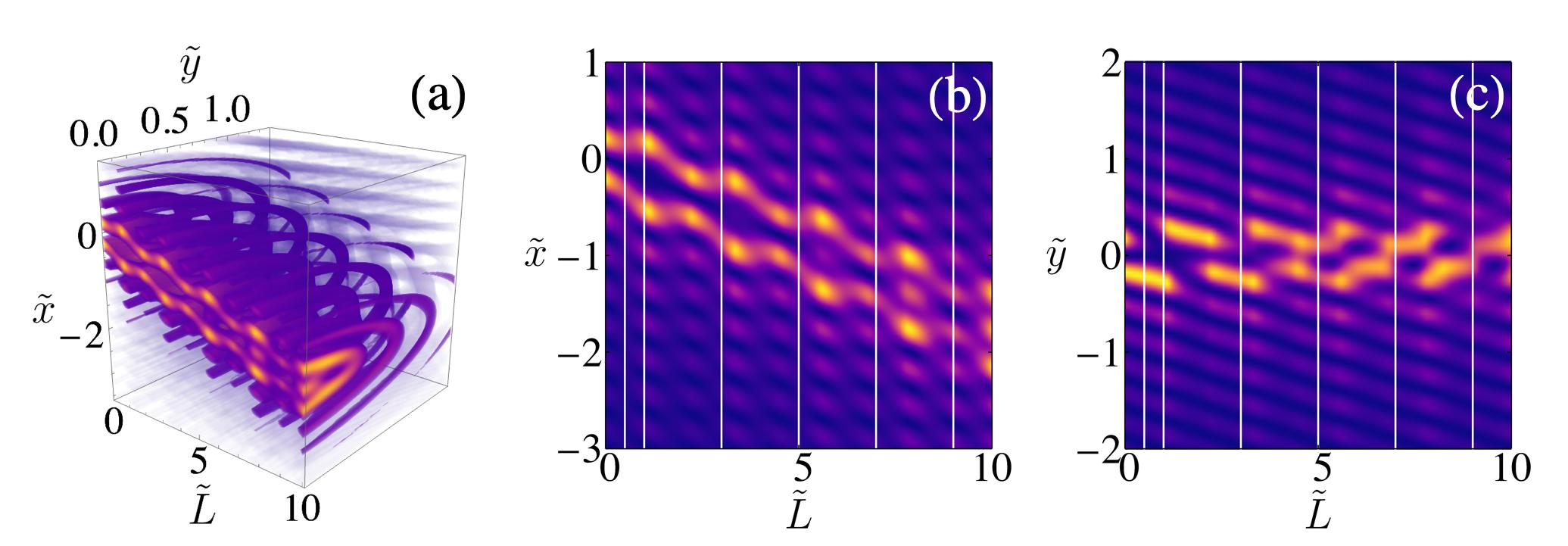}
\end{center}
\caption{ {\bf Propagation of the intensity $|\mathbf{E}^{\mathbf{in}}|^2$ and its projections for an incident circularly polarized Bessel beam }. (a) Propagation along the crystal length $\tilde{L}=L\times 10^{-3}$, $\tilde{x}=x\times10^{-2}$ and $\tilde{y}=y\times10^{-2}$. (b) Projection of the propagation for $\tilde{y}=0$. (c) Projection of the propagation for $\tilde{x}=-1.5\tilde{ L}\times 10^{-1}$, the approximate axis for the maxima in the intensity profile. Crystal parameters are the same as in figure \ref{Fig2}.}
\label{Fig9}
\end{figure*}

\begin{figure*}[htbp!]
\begin{center}
\includegraphics[width=0.5\textwidth]{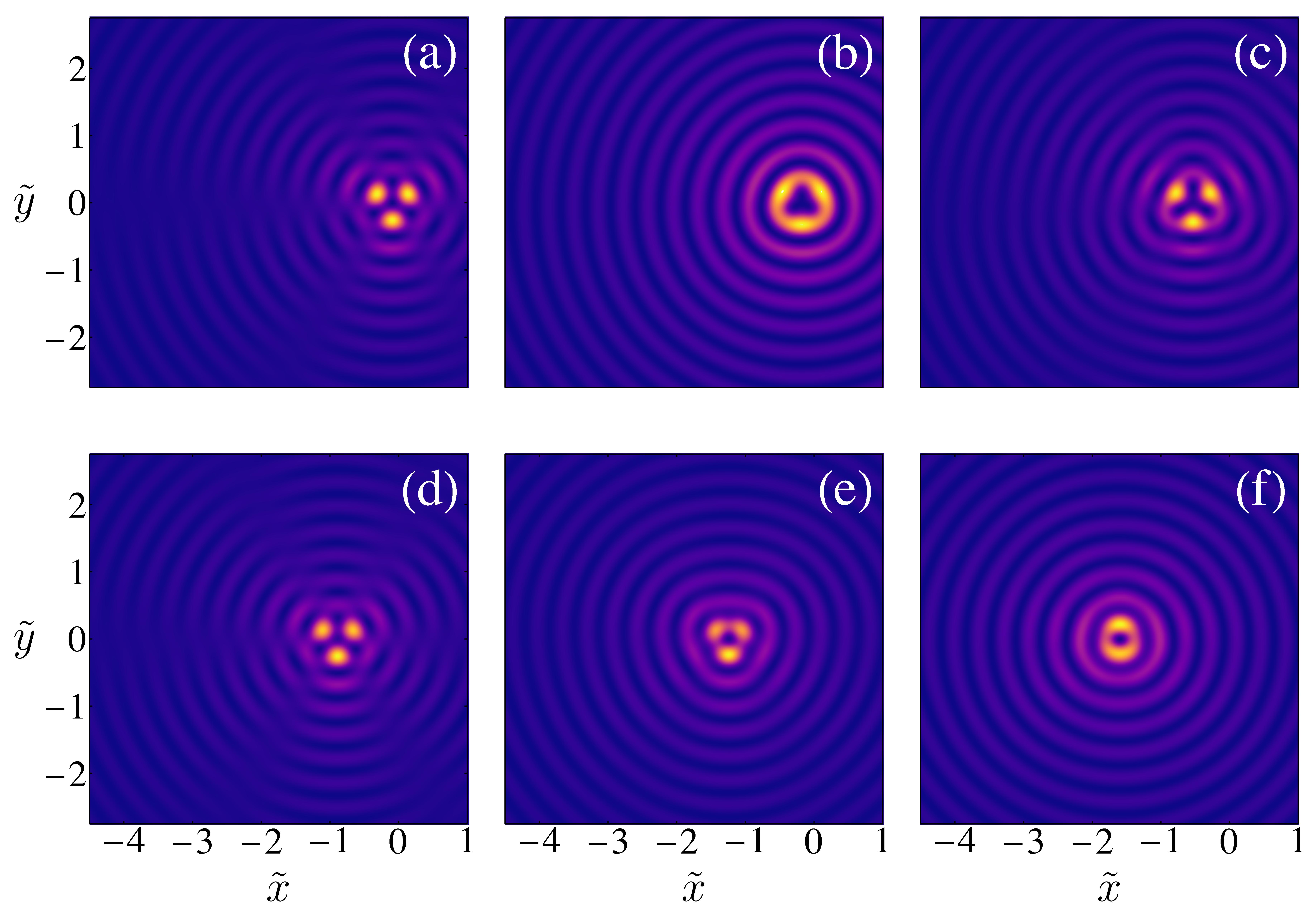}
\end{center}
\caption{ {\bf Transverse planes at different propagation distances $\tilde{L}$ for or an incident circularly polarized Bessel beam}. The distance propagated along the crystal corresponds to the white lines in  Fig. \ref{Fig9} (b) and (c). The distances are $\tilde{L}=$ 0.5 (a) , 1  (b), 3 (c) , 5 (d), 7 (e), 9 (f). Parameters are the same as in figure \ref{Fig9}.
 }
 \label{Fig10}
\end{figure*}

We integrate numerically by standard methods, using the Simpson's rule \cite{NumericalRecipes}. In general, it is convenient to perform the integration in the $k_\perp$ and $\phi'$ variables. We implement the integration subroutine and solution of the systems \eqref{111} and \eqref{222} using a multithreaded code implemented in C++ in the case where angular integration is only needed, as for a Bessel incident beam. However, when integrals involve both $k_\perp$ and $\phi'$, the computational times increase dramatically even for multi-threaded implementations. To circumvent this, the numerical integration code was implemented using C++ with CUDA extensions\cite{CUDA} and it was run in Nvidia GPU's. The use of the GPU's substantially improved the computational times, reducing them several orders of magnitude from projected calculated times of weeks to minutes. This allowed to arbitrarily simulate the propagation to very long distances $L=10^4$ with high numerical accuracy and very small grid spacing in the integrations. All the numerical simulations have machine precision error and for practical purposes are numerically exact. Simulations were run in a server with an Epyc AMD dual socket CPU with 96 cores and 2 Nvidia T4 GPU accelerators part of the LSCSC-LANMAC infrastructure.  Results of the numerical simulations are presented in figures \ref{Fig3},\ref{Fig4},\ref{Fig5},\ref{Fig6},\ref{Fig7},\ref{Fig8},\ref{Fig9}, and \ref{Fig10}. Note that in these figures we have normalized  the intensity $|\mathbf{E}^{\mathbf{in}}|^2$  with respect to its maximum value at each $\tilde{L}$. Typical parameters of the simulations for the integration in the Bessel case are grids of 1024 to 4096 points, and for the Gaussian case grids of 256 to 1024 points in $\phi'$ and 1024 to 4096 points in $k_\perp$. 

In figures \ref{Fig3} and \ref{Fig4}, we show the propagation inside the crystal of a Gaussian  beam incident in the $\hat{e}_x$ direction. As shown in the scheme of figure \ref{Fig1}. We find that the diffraction cone opens as the beam propagates inside the crystal. The cone opens asymmetrically, as shown in the transverse planes at different crystal lengths,  figure \ref{Fig4}. In contrast to this, when the beam is circularly polarized in figures ref{Fig5} and \ref{Fig6}, we find that the cone is symmetric. The reason for this is that both polarizations in the $\hat{e}_x$ and $\hat{e}_y$ are balanced. Thus, as one changes the proportion between polarizations, one can go from an asymmetric cone in the $\hat{e}_x$ axis to a symmetric one in the circularly polarized case. This process is symmetrical with respect to the change of the initial polarization axis to $\hat{e}_y$.

For the profile of the incident Bessel beam, we consider the linearly polarized case in figures \ref{Fig7} and \ref{Fig8} and the circularly polarized case in  figures  \ref{Fig9} and \ref{Fig10}. Here, in contrast with the gaussian cases, we find that the diffraction cone does not occur. This is due to the property of Bessel beams of being diffrection free \cite{Everly}, and it could have been expected since we are considering a linear though birefringent medium. However, we find that there are formations of  regions of minimal intensity in the center of the propagated beams.  Interestingly, we find that the beam propagated in the crystal mixes several components of higher order Bessel functions, similar to what was reported in \cite{King}.  This leads to the formation of maxima around the dark region in the center of the intensity profile that rotates and mixes as the beam propagates, see  Figs. \ref{Fig8}  and \ref{Fig10}. While a Bessel beam does not form a diffraction cone, we find that the beam gets deflected  approximately following the directrix of the diffraction cone, but at a smaller slope than that of  the gaussian case.  The effect of the different chosen polarizations is that, for the linear case, one can observe that there are regions where the maxima in the center of the beam get strongly suppressed, with dark regions as in figure \ref{Fig7}(a) and (c). In contrast to this,  the maxima are approximately constant in the circularly polarized case, see figure \ref{Fig9}(a) and (c). We verified this fact changing from right to left circularly polarized beams and we found that the results are essentially the same up to a rotation of $90^\circ$ in the $x-y$ plane.

\section{Description of the numerical scheme used for the simulations}
The steps we follow, given the parameters of the KTP crystal are:
\begin{itemize}
\item{We find numerically the real positive solutions \eqref{F2}, using standard methods, i.e. Newton-Raphson~\cite{NumericalRecipes}}
\item{We generate a high order interpolation polynomial (IP) with the solution of \eqref{F2} for $K_\pm$}
\item{With the IP, we construct the system of 12 equations  (real and imaginary parts) given by \eqref{111} and \eqref{222}, given an incident electric field profile in position space for each cartesian point at a crystal length $L_n$.}
\item{We numerically solve the system of equations using standard Linear Algebra subroutines\cite{NumericalRecipes}}
\item{With the solution of $\tilde{E}^{\pm}_{x,y,z}$ we integrate over momentum space (using Simpson's rule), in the $k_x-k_y$ plane or for fixed $k_\perp$ for the Bessel incident beams.}
\item{We change the crystal length $L_n\to L_{n+1}$ and repeat until we reach the desired length of the crystal $L$.}
\end{itemize}

As the algorithm is not dependent on previous steps in the propagation inside the crystal, therefore it can be fully parallelized. 
\section{Conclusions}

Our methods and simulations can be extended to arbitrary incident profiles and linear crystals with more elaborated tensor parameters and less symmetry. Possible extensions of our methods include the analysis of propagation in nonlinear media and analogous systems, such as cold matter \cite{Daniel}. In any case, it is clear from our numerical results that the phenomenon of conic refraction is very sensitive to the initial conditions provided by the impinging beam on the crystal. Our study suggests that in practice a Gaussian beam is the best option for producing this very special effect in a laboratory.

{\bf Aknowledgements.} This work was partially supported by the grants UNAM, DGAPA-PAPIIT:  IN109619, UNAM-AG810720, LANMAC-2019 and CONACYT Ciencia B\'asica: A1-S-30934. 
We acknowledge infrastructure support for the computations from the ``Laboratorio de Simulaciones Computacionales para Sistemas Cu\'anticos'' in LANMAC (LSCSC-LANMAC) at IF-UNAM. 

{\bf Disclosures.} The authors declare no conflicts of interest.

\end{document}